\documentclass[12pt,preprint]{aastex}

\shorttitle{Two blue-straggler sequences}
\shortauthors{Jiang et al.}

\begin{document}



\title{Contribution of primordial binary evolution to the two
blue-straggler sequences in globular cluster M30}

\author{Dengkai Jiang\altaffilmark{1,2,3}, Xuefei Chen\altaffilmark{1,2,3}, Lifang Li\altaffilmark{1,2,3} and Zhanwen
Han\altaffilmark{1,2,3}}

\altaffiltext{1}{Yunnan Observatories, Chinese Academy of
Sciences, 396
 Yangfangwang, Guandu District, Kunming, 650216, P.R. China; dengkai@ynao.ac.cn, zhanwenhan@ynao.ac.cn}

\altaffiltext{2}{Center for Astronomical Mega-Science, Chinese
Academy of Sciences,
  20A Datun Road, Chaoyang District, Beijing, 100012, P.R. China}

\altaffiltext{3}{Key Laboratory for the Structure and Evolution
of Celestial Objects, Chinese Academy of Sciences, Kunming,
650011, China}

\begin{abstract}

Two blue-straggler sequences discovered in globular cluster M30
provide a strong constraint on the formation mechanisms of blue
stragglers. We study the formation of the blue-straggler
binaries through binary evolution, and find that binary
evolution can contribute to the blue stragglers in both of the
sequences. Whether a blue straggler is located in the blue
sequence or red sequence depends on the contribution of the
mass donor to the total luminosity of the binary, which
is generally observed as a single star in globular clusters.
The blue stragglers in the blue sequence have a cool
white-dwarf companion, while the majority ($\sim 60\%$)
of the objects in the red sequence are the binaries that are
still experiencing mass transfer, but there are also some
objects that the donors have just finished the mass transfer
(the stripped-core stars, $\sim 10\%$) or the blue
stragglers (the accretors) have evolved away from the blue
sequence ($\sim 30\%$). Meanwhile, W UMa contact binaries
found in both sequences may be explained by various mass
ratios, that is, W UMa contact binaries in the red sequence
have two components with comparable masses (e.g. mass ratio
$q\sim$ 0.3-1.0), while those in the blue sequence have low
mass ratio (e.g. $q<$ 0.3). However, the fraction of blue
sequence in M30 cannot be reproduced by binary population
synthesis if we assumed the initial parameters of a binary
sample to be the same as that of the field. This possible
indicates that dynamical effects on binary systems is very
important in globular clusters.

\end{abstract}


\section{Introduction}

Blue stragglers are a class of anomalous stars that are
brighter and bluer than the main-sequence (MS) turnoff stars in
the color-magnitude diagram of globular clusters. They are very
common objects in almost all Galactic globular clusters
\citep{Piotto2004}, and can be used to probe the dynamical
evolution of clusters \citep{Ferraro2012}. Their locations in
the color-magnitude diagram suggests that they may be MS stars
more massive than typical MS turnoff stars \citep{Ferraro2006},
and they should have evolved away from the main sequence.

At present, there are two popular mechanisms to explain the
formation of blue stragglers: binary evolution
\citep{McCrea1964} and direct stellar collision
\citep{Hills1976}, and a series of work have been done for the
two mechanisms in recent ten years
\citep[e.g.][]{Sills1999,Sills2000,Xin2007,Tian2006,Chen2008,Chen2008b,Chen2009,Lu2010,Leigh2013}.
It is generally believed that binary evolution plays an
important role in open clusters and in the field, while direct
stellar collision are likely important in dense environments
such as globular clusters or the core of open clusters
\citep{Hills1976, Sills2002,
Glebbeek2008,Mathieu2009,Geller2011}. However, observations
show that the two mechanisms may be important in the same
clusters
\citep{Ferraro1993,Ferraro1995,Ferraro1997,Ferraro2004}.

An important and perhaps critical clue to the origin of blue
stragglers is the two blue-straggler sequences observed in the
color-magnitude diagram of globular cluster M30
\citep{Ferraro2009}. Similar features are also found in NGC 362
\citep{Dalessandro2013} and NGC 1261 \citep{Simunovic2014}. The
occurrence of two sequences can be explained by the coexistence
of blue stragglers formed through two different formation
mechanisms enhanced by core collapse 1-2 Gyr ago
\citep{Ferraro2009}. Each of the two sequences may correspond
to a distinct formation mechanism \citep{Ferraro2009}, because
the blue sequence is outside the ``low-luminosity boundary"
defined by the binaries with ongoing mass transfer
\citep{Tian2006,Xin2015} and the red one is too red to be
reproduced by collisional models \citep{Sills2009}. However,
NGC 1261, one of three globular clusters with two
blue-straggler sequences, does not show the classical
signatures of core-collapse \citep{Simunovic2014}.

It should be noted that three W UMa contact binaries have been
detected in both sequences of blue stragglers in M30
\citep{Pietrukowicz2004, Ferraro2009}. W UMa contact binaries
are very common among blue stragglers in globular clusters
\citep{Rucinski2000}, which are thought to come mainly from
binary evolution \citep{Vilhu1982,Jiang2014a}. Hence, the
formation of both sequences may be related to the binary
evolution. Meanwhile, \citet{Lu2010} found that some blue
stragglers produced by Case B binary evolution are below the
low-luminosity boundary given by \citet{Tian2006}.
\citet{Chen2008} show that binary merger can produce single
blue stragglers very close to or even below the zero-age main
sequence (ZAMS), i.e. in the blue sequence. In addition,
\citet{Stepien2015} found that binary merger can form blue
sequence of blue stragglers while binary blue stragglers can
lead to a red sequence. Therefore, more study about the
formation of blue stragglers by binary evolution should be done
to check whether binary evolution can provide a contribution to
the formation of the blue-straggler blue sequence in globular
cluster M30.

\section{The possibility of binary evolution contributing to
the blue-sequence blue stragglers}

Before detailed binary evolution calculations are performed, we
will simply discuss the possibility of binary evolution
contributing to the blue-sequence blue stragglers. At first, if
not considering contact binaries, binary evolution can produce
two kinds of blue-straggler binaries (as shown in Figure 1):
those are still experiencing mass transfer \citep[e.g. V228 in
47 Tuc,][]{Kaluzny2007} and those have finished mass transfer
\citep[e.g. WOCS 4348, 4540 and 5379 in NGC
188,][]{Gosnell2014}. The blue-straggler binaries during the
mass-transfer phase have a ``low-luminosity boundary" (about
0.75\,mag brighter than the ZAMS) given by \citet{Tian2006},
and can match the observed red-sequence blue stragglers in
globular cluster M30 \citep{Ferraro2009, Xin2015}. However, for
the blue-straggler binaries that have finished mass transfer,
including a blue straggler and a white dwarf (the BS-WD
binaries), their locations in the color-magnitude diagram of
M30 depend on the contribution of white dwarfs to the
combined magnitudes of these binaries.

We can simply estimate the location of a BS-WD binary in the
color-magnitude diagram as follows. We take the binaries with a
0.8\,M$_{\rm \odot}$ primary (metallicity $Z = 0.0003$) as
examples, and the secondary masses are taken to be 0.75, 0.7,
0.65,..., 0.3M$_{\rm \odot}$. These binaries are assumed to
experience Case B mass transfer at about 12\,Gyr. The primaries
will transfer their envelopes (about 0.55 M$_{\rm \odot}$ in
conservative case of mass transfer) to the secondaries, leaving
a helium WD star (about 0.25M$_{\rm \odot}$). The secondaries
that gain mass will rejuvenate and evolve up along the main
sequence to higher luminosity and effective temperature
\citep{Tout1997, Hurley2002}. When mass transfer finished
(assuming at about 12.5\,Gyr), the secondaries become blue
stragglers with masses of 1.3, 1.25, 1.2,..., 0.85\,M$_{\rm
\odot}$ as rejuvenated star. We approximate this
rejuvenation\footnote{According to the description of
\citet{Tout1997}, the rejuvenation of main-sequence stars with
no convective core (0.3 $\sim$ 1.3M$_{\rm \odot}$) can be
approximated by taking the remaining fraction of main-sequence
life directly proportional to the remaining fraction of unburnt
hydrogen at the centre, and adjusting the effective age $t$ of
the stars, $t' = t \times ( \tau'_{\rm MS}/\tau_{\rm MS})$. }
as described by \citet{Tout1997} and \citet{Hurley2002}. After
the rejuvenation, these blue stragglers continue to evolve as
single stars.

At the time of the end of mass transfer, the primaries are the
stripped giant stars, which are brighter and redder than the
turnoff. So we simply assume that they have the same magnitude
and color as that of a single star with 0.8\,M$_{\rm \odot}$ at
the red giant branch (e.g. $V \sim 2.5$ and $V-I \sim 0.6$). By
combining their companions (the rejuvenated stars), the
combined magnitude of these BS-WD binaries can be calculated
using a formula given by \citet{Xin2015}, for example, the
V-band magnitude of the binary system
\begin{equation}
V = V_{\rm 1}-2.5 {\rm log} (1+10^{(V_{\rm 1}-V_{\rm
2})/2.5}),
\end{equation}
where $V_{\rm 1}$ and $V_{\rm 2}$ are the V-band magnitudes of
two components, respectively. After mass transfer terminates,
these primaries evolve quickly to a helium white dwarf and cool
down. According to the equation of luminosity evolution of
white dwarfs given by \citet{Hurley2000},
\begin{equation}
L_{\rm WD} = \frac{635MZ^{0.4}}{[(A(t+0.1))]^{1.4}}
\end{equation}
(where $M, Z, A$ and $t$ are the mass, metallicity,
effective baryon number and age of white dwarfs,
respectively), these white dwarfs would have much lower
luminosities than the blue-straggler companion when they cool
to the age of M30 (13\,Gyr) from the end of mass transfer
(12.5\,Gyr). Here, we roughly assume that the $V$-band
magnitude of the white dwarfs increase with their age (e.g.
$V_{\rm WD}=V_{\rm BS}+3$ at 13.0\,Gyr, $V_{\rm WD}=V_{\rm
BS}+4$ at 14.0\,Gyr), while these white dwarfs have a color
$V-I=-0.2$.

In figure 2a, we show the locations of these BS-WD binaries in
the color-magnitude diagram at the time of the end of mass
transfer(12.5\,Gyr), at the age of M30 (13.0\,Gyr), and at
their subsequent evolution (e.g. 14.0\,Gyr). When the mass
transfer terminates, these binaries are above the
``low-luminosity boundary", which is 0.75\,mag brighter than
the ZMAS as suggested by \citet{Tian2006}. They then move to
the region below the ``low-luminosity boundary" when the
primaries become a helium WD and cool to the age of M30. They
mainly appear in the location between the ZAMS and the boundary
at 13.0\,Gyr and 14.0\,Gyr, where the blue sequence defined by
\citet{Ferraro2009}. Therefore, it is possible that the BS-WD
binaries may contribute to the blue sequence in M30.

It should be noted that at the time of the end of mass
transfer, these binaries have very different locations from the
blue-straggler components as shown in Figure 2a, because the
primaries (stripped giant stars) are brighter than the
blue-straggler components, dominating the positions of these
binaries. As the white dwarfs cool, they are closer to the
BS-WD binaries, and almost overlap the BS-WD binaries at
14.0\,Gyr. The faint blue-straggler components are close to the
ZAMS because their progenitors are less evolved, low-mass
stars. Considering the effect of non-conservative mass transfer
(e.g. 50\% of the transferred mass is assumed to be lost during
mass transfer), the BS-WD binaries are still below the
``low-luminosity boundary". However, there are no bright and
blue BS-WD binaries because of the decrease of accreted mass of
the secondaries. Moreover, in Figure 2b, we compare these BS-WD
binaries with the collision isochrones corresponding to ages of
1 and 2\,Gyr, which agree with the observed blue sequence as
shown in Figure 4 of \citet{Ferraro2009} and given by
\citet{Sills2009}. These collisional isochrones have been
transferred into the absolute plane using a distance modulus of
$(m-M)_{\rm v} = 15.04$\,mag and a reddening of $E(V-I) =
0.112$\,mag, and the values of distance modulus and reddening
are obtained by comparing the location of our 13\,Gyr isochrone
of single stars and ZAMS (the blue dotted and blue dashed
lines) with those lines in \citet{Ferraro2009} (the red dotted
and red dashed lines). As shown in Figure 2b, the BS-WD
binaries are in the similar region to the collision isochrones,
which are between the ZAMS and the ``low-luminosity boundary".

\section{Binary evolution calculations}
We have carried out a detailed study of the formation of blue
stragglers from binary evolution using Eggleton's stellar
evolution code. This code is a variant of the code \textbf{ev}
described, in its initial version, by
\citet{Eggleton1971,Eggleton1972} and \citet{Eggleton1973a},
updated during the last four decades
\citep{Eggleton1973b,Han1994,Pols1995,Nelson2001,Eggleton2002,Yakut2005,Eggleton2006}.
The current version of \textbf{ev} (private communication 2003)
is obtainable on request from Peter.Eggleton@yahoo.com, along
with data files and a user manual. We calculate the evolution
of binaries with metallicity $Z = 0.0003$, which are close to
the metallicity ([Fe/H]$=-1.9$) of M30 \citep{Ferraro2009}. We
construct a grid of conservative binary evolutionary models
from ZAMS to the age of M30 (13\,Gyr), with the following
ranges of initial primary mass $M_{10}$, initial mass ratio
($q_0=M_{20}/M_{10}$) and initial orbital period $P_{0}$:
\begin{equation}
\log M_{10} = -0.110, -0.105, -0.100,..., -0.02, \,
\end{equation}
\begin{equation}
\log (1/q_{0}) = 0.025, 0.050, 0.075,..., 0.600, \,
\end{equation}
\begin{equation}
\log (P_0/P_{\rm ZAMS}) = 0.025, 0.050, 0.075,..., 1.000, \,
\end{equation}
where $P_{\rm ZAMS}$ is the period at which the primary would
just fill its Roche lobe on the ZAMS \citep{Nelson2001}. We
assume that the binary orbit is circular because it will
circularize quickly during the mass transfer\footnote{The
originally eccentric orbits would be circularized quickly
during mass transfer \citep{Hurley2002}, and only a few evolved
binaries have eccentric orbits \citep[e.g.][]{deMink2007}.
However, the parameter space for the formation of blue
stragglers at 13\,Gyr may be smaller if the mass transfer will
not always lead to circular orbits as suggested by
\citet{Sepinsky2007,Sepinsky2009,Sepinsky2010}. This is because
mass transfer in the eccentric binaries may be episodic (only
at periastron), which results in the less-massive, more evolved
accretor stars than that in the circular binaries.}. In
Eggleton's Stellar evolution code, the magnitude of each
component is given from luminosity and effective temperature
based on the table given by \citet{Flower1996}. Then we
calculate the combined magnitude of blue-straggler
binaries using equation (1).

Four representative examples of our binary evolution
calculations are shown in Figure 3. The first example is a
blue-straggler binary that is experiencing mass transfer, and
appears in the red sequence at the age of M30. The other three
examples are the BS-WD binaries that can evolve into the blue
sequence. However, at the age of M30, the second example
appears in the blue sequence, while the third and fourth
examples appear in the red sequence.

Figures 3a and 3b show the evolutionary track of the first
example, including the combined evolutionary tracks of
the binary system and both components. Mass transfer begins at
10.44\,Gyr, and this system evolves into the region of blue
straggler along a line parallel to the ``low-luminosity
boundary". This binary is in the observed red-sequence region
at 13\,Gyr, while it is still in the mass-transfer stage. At
this time, the donor star is in the red giant branch at
13\,Gyr, and its luminosity is high enough to significantly
change the position of this binary relative to the accretor
star ($\bigtriangleup$$V = -0.2$; $\bigtriangleup$$(V-I) =
-0.14$) in the color-magnitude diagram. Finally, this binary
leaves the blue-straggler region at about 13.63\,Gyr.

For the second example in Figures 3c and 3d, Mass transfer
between two stars begins and terminates at 9.91 and 10.86\,Gyr,
respectively, and then this BS-WD binary evolves across the
``low-luminosity boundary". At the age of globular cluster M30
(13\,Gyr), this binary is in the region between the ZAMS and
the ``low-luminosity boundary", where the observed blue
sequence in M30 lies. It should be noted that after the end of
mass transfer, the location of this BS-WD binary depends on two
timescales: (1) the timescale for the white dwarf that cools to
almost no contribution to $V$-band magnitude (e.g. the location
of the BS-WD binary having largest $V$-band magnitude at
11.08\,Gyr), which is about 0.22\,Gyr; (2) the timescale for
the remaining main-sequence lifetime of the blue straggler
(about 2\,Gyr). The cooling timescale is much shorter than the
remaining MS lifetime of the blue straggler. Therefore, this
BS-WD binary can appear below the ``low-luminosity boundary",
like a``single" blue straggler. Although the third and fourth
examples (Figure 3e and 3g) are BS-WD binaries that have
similar evolutionary tracks to the second example, they appear
above the ``low-luminosity boundary" in the color-magnitude
diagram at the age of M30, before they evolve into the blue
sequence or after they evolve away the blue sequence.

The products of binary evolution, the blue-straggler binaries
in the two sequences, may have different initial binary
parameters. We show their initial binary parameters in the
initial orbital period-secondary mass planes for different
initial primary mass (Figure 4). They are roughly classified
based on above or below the ``low-luminosity boundary"
(0.75\,mag brighter than the ZAMS). The progenitors of the blue
sequence are constrained to host 0.80-0.93\,M$_{\rm \odot}$
primaries and 0.23-0.76\,M$_{\rm \odot}$ secondaries in orbits
with initial orbital periods of 0.4-2.3\,d. The range of
initial primary mass for the red sequence is slightly larger
than that for the blue sequence. However, the blue stragglers
in the red sequence have a shorter initial orbital period and a
more massive initial secondary than those in the blue sequence
with the same initial primary masses. In general, the blue
sequence mainly comes from case B binary evolution, while the
red sequence mainly comes from case A binary evolution.

\section{Binary population synthesis}

In order to estimate the distribution of blue-straggler
binaries, we performed a series of Monte Carlo simulations (see
Table 1) based on our grid of conservative binary evolutionary
models described above. The following input is adopted for the
simulation \citep{Han1995}. (1) The initial mass function (IMF)
of \citet[][MS79]{Miller1979} is adopted. An alternative IMF of
\citet[][S86]{Scalo1986} is also considered. (2) We adopt three
initial mass-ratio distributions: a constant mass-ratio
distribution, a rising mass-ratio distribution and one where
the component masses are uncorrelated. (3) The distribution of
separations is taken to be constant in log\,$a$ for wide
binaries, where $a$ is the orbital separation. We also take a
different period distribution of \citet[][DM91]{Duquennoy1991}.
(4) A circular orbit is assumed for all binaries.

Figure 5 shows the result of the simulation set 1 at 13\,Gyr in
the color-magnitude diagram. It is striking that this
simulation agrees quite well with the distribution of the
observed blue stragglers in M30. There is a blue-straggler
sequence, similar to the observed blue sequence, which is
between the ZAMS and the ``low-luminosity boundary" and about
half a magnitude brighter than the ZAMS. Meanwhile, the blue
stragglers above the "low-luminosity boundary" cover a wider,
sparser area, which agrees with the distribution of the
observed red sequence. The blue-sequence binaries have a blue
straggler orbiting a white dwarf, while those red-sequence
binaries include the binaries that are experiencing mass
transfer ($\sim $60\%), or just terminate mass transfer ($\sim
$10\%), and the binaries that the blue stragglers have evolved
away from the blue sequence ($\sim $30\%). The results of
other Monte Carlo simulations (sets 2 to 5) are plotted in
Figure 6, and these simulations give similar results to the
simulation set 1. These four simulations also show the presence
of two blue-straggler sequences that are in agreement with the
observed distribution of blue stragglers in M30.

To estimate the total number of blue stragglers and the
fraction of blue sequence in M30 from binary population
synthesis, we assume an initial binary fraction ($f_{\rm b}$)
of M30, e.g. 25\%, which is half of the binary fraction in the
solar neighbourhood \citep[50\%,][]{Halbwachs2003}. As
alternatives we also consider a binary fraction, 15\%. These
fractions are assumed to be higher than the binary fraction of
M30 at present day \citep[about 7\%,][]{Milone2012} because the
binary fraction decreases with time due to the dynamical
interactions and binary evolution \citep{Ivanova2005}. In
addition, the initial mass of M30 is simply assumed to be twice
as massive as the current mass of M30 \citep[log\,(M/M$_{\rm
\odot}$)=5.3,][]{Sandquist1999} as the globular clusters may
have lost a significant fraction of total mass driven by
relaxation, stellar evolution and the tidal field of the Galaxy
\citep{Vesperini1997}. Because the binaries are more difficult
to lose from the globular clusters than single stars, the
binary fraction in the lost stars is assumed to be half of the
initial binary fraction.

The results are summarized in Table 2 for the total numbers of
blue stragglers ($N_{\rm total}$) and the fraction of blue
sequence ($N_{\rm blue}$/$N_{\rm total}$). The total numbers of
blue stragglers range from 10 to 99, and the fraction of blue
sequence ranges from 60\% to 70\%. It is clear that the initial
distributions of binaries are very important in determining the
formation of blue stragglers from binary evolution. The
uncorrelated mass-ratio distribution (set 3) or the nonconstant
distribution of orbital separation (set 5) makes a smaller
$N_{\rm total}$ and a larger $N_{\rm blue}$/$N_{\rm total}$, as
compared to the simulation set 1. On the other hand, the rising
mass-ratio distribution (set 2) or the IMF of of
\citet{Scalo1986} makes a larger $N_{\rm total}$ and a smaller
$N_{\rm blue}$/$N_{\rm total}$. The total number of blue
stragglers depends strongly on $f_{\rm b}$, although the
fraction of blue sequence does not depend on $f_{\rm b}$. Based
on the observed results given by \citet{Ferraro2009} and
\citet{Xin2015} (as shown in Figure 5), the observed values of
$N_{\rm total}$ and $N_{\rm blue}$/$N_{\rm total}$ are 49 and
49\%, respectively (25 red-sequence stars and 24 blue-sequence
stars). The results of binary population synthesis can explain
the observed total numbers of blue stragglers in M30, but fail
to explain the fraction of blue sequence in M30.

\section{Discussion}
\subsection{The fraction of blue sequence}

In our study, the fraction of blue sequence cannot be
reproduced by binary population synthesis. Our simulations
predict that $60\%-70\%$ of the total blue stragglers should be
observed in the blue sequence, while the observed blue sequence
only contains 49\% of the total blue stragglers in
M30\footnote{If four red-sequence stars below the
``low-luminosity boundary" (red points with blue circles in
Figure 5) are classified as blue sequence, the observed
fraction of blue sequence is 57\%.}. If the contribution of
direct stellar collision and binary merger to the blue sequence
is considered, the problem is more challenging.

One possible explanation is mass loss during mass transfer
which has to be considered in our present study, and we would
overestimate the fraction of blue stragglers in the blue
sequence. As examples shown in Figure 7, we compare the
relative regions of the progenitors for two sequences in the
conservative and non-conservative cases (e.g. 50\% of the
transferred mass is assumed to be lost from the system). We
find that for the non-conservative cases, the progenitor region
of the blue sequence is reduced more remarkably, relative to
the region of the red sequence. But we should note that the
brightest blue stragglers are significantly fainter that shown
in previous study for the non-conservative assumption since the
accretor cannot increase mass as much as that. This may
decrease the percentage of the blue sequence in the binary
scenario.

Another possible explanation is the uncertainties in the
distribution of binary parameters in globular clusters that are
important to binary population synthesis. The uncertainties do
not change the appearance of two blue-straggler sequences as
shown in Figure 6, but significantly alter the quantitative
estimates of the total number of blue stragglers and the
fraction of blue sequence. At present, our results of binary
population synthesis are based on the assumptions that adopted
in the field. It is very likely that dynamical interactions in
globular clusters alter the parameter distribution of
primordial binary population. For example, the Heggie-Hills law
\citep[hard binaries get harder,][]{Heggie1975, Hills1975}
tends to make the binaries closer, and exchange encounters
(often eject the least massive of the three stars) are more
likely to increase the mass ratio of binaries. Based on the
initial distribution of binaries shown in Figure 4, these
dynamical effects can decrease the fraction of blue sequence by
bringing the binary systems from Case B evolution to Case A
evolution. Therefore, it is very important to understand these
uncertainties in the distribution of binaries in globular
clusters.

Our results do not rule out the contribution of dynamical
interactions, especially the core collapse, to the formation of
two blue-straggler sequences in M30. Our results show that
binary evolution can produce a blue sequence below the
"low-luminosity boundary", but this sequence is not as tight as
the blue sequence observed in M30. This tight blue sequence may
come from core collapse \citep{Ferraro2009}, which limited the
time range for the formation of blue-sequence blue stragglers
from binary evolution and direct stellar collision.

From present results of binary evolution, we predict that the
majority of binary-origin blue stragglers in the blue sequence
should have a low-luminosity white-dwarf companion if they are
not already disrupted due to dynamical interactions. Meantime,
not all blue stragglers in the red sequence are experiencing
mass transfer, and some of them may also have a white-dwarf
companion. Moreover, the blue sequence may show chemical
anomalies (as a significant depletion of carbon and oxygen),
similar to the red sequence with O-depletion \citep{Lovisi2013}
because chemical anomalies are expected for the binary-origin
blue stragglers \citep{Chen2004,Ferraro2006,Jiang2014b}, but
not for the collision-origin blue stragglers
\citep{Lombardi1995}. Future observations of two sequences of
blue stragglers would determine whether binary evolution can
contribute to both sequences of blue stragglers in M30.

\subsection{Comparisons with previous studies}

Our results are consistent with previous theoretical studies.
It has been indicated that binary evolution can produce blue
stragglers below the ``low-luminosity boundary" from binary
merger \citep{Chen2008, Stepien2015} and case B mass transfer
\citep{Lu2010}. \citet{Lu2010} have shown that the case B
binary evolution can reproduce a bluer sequence of blue
stragglers than those from case A binary evolution. Moreover,
the observations have shown that in the color-magnitude diagram
of open cluster NGC 188, some binary blue stragglers are close
to the ZAMS as shown in Figure 1 of \citet{Mathieu2009}. So far
these observed blue stragglers have been interpreted to have a
binary origin with white dwarf companions
\citep{Geller2011,Gosnell2015}, which are similar to those
BS-WD binaries in our models.

\citet{Xin2015} investigated the binary origin of blue
stragglers in M30 by using a different version of Eggleton's
stellar evolution code. They showed that the binary models
nicely match the observed red sequence in M30, but can not
attain the observed location of blue sequence. Their
calculations missed those binary models that can produce blue
stragglers in the blue sequence, maybe because their grid
covers a larger range of primary from 0.7\,M$_{\rm \odot}$ to
1.1\,M$_{\rm \odot}$, but with larger steps of 0.1\,M$_{\rm
\odot}$. Moreover, their code stopped in some cases because of
numerical instabilities that prevent to follow the complete
evolutionary tracks of these systems \citep{Xin2015}.

\subsection{Special or common phenomenon in globular clusters?}
We suggest that the age of M30 (13\,Gyr) is not special for the
formation of two blue-straggler sequences in globular clusters
from binary evolution, because the accretor stars with a white
dwarf become inevitable when the mass transfer is finished. For
example, a binary system with $M_{10}=1.0$\,M$_{\rm \odot}$
($M_{20}=0.45$\,M$_{\rm \odot}$, $P_0=0.677$\,d) can be located
in the blue sequence at 8\,Gyr as shown in Figure 8, which
suggests that the different primary mass of the binary may
contribute to the blue stragglers in the blue sequence at
various age. Therefore, the appearance of two sequences
produced by binary evolution may not be a short-lived
phenomenon in globular clusters.

However, clearly separated sequences are not easy to be
observed. It should be noted that in the other two globular
clusters (NGC 362 and NGC 1261), the gaps between two sequences
are much smaller than the widths of blue sequence. This may be
because of reddening variation, distance variation,
observational error etc. For example, the core of globular
clusters may be the most probable place showing two clearly
separated sequences, in which all blue stragglers can be
thought to have the exact same distance and reddening, without
the pollution from the blue stragglers in the outer region of
cluster that may have a slightly different reddening and
distance modulus. Moreover, the photometric error needs to be
significantly smaller than the shift by the companions in the
color-magnitude diagram. Considering the subsequent evolution
of blue-sequence blue stragglers, some of them may also appear
between two sequences, e.g. the brightest blue-sequence blue
straggler observed in M30 that is very close the
``low-luminosity boundary" as shown in Figure 5.

\subsection{W UMa contact binaries in the blue-straggler region}

Our simulations cannot take into account W UMa contact binaries
because of the numerical difficulty of constructing their
physical models, which is still one of the most important
unsolved problems of stellar evolution
\citep{Eggleton2006a,deMink2007,Eggleton2010}. However,
\citet{Rucinski2004} has shown that {the total luminosity} of W
UMa contact binaries becomes more and more similar to that of
the bright component in such a binary with the mass ratio
decreasing, since the contribution from the faint one becomes
smaller and smaller. Here we estimate roughly their locations
in the color-magnitude diagram in a similar way to
\citet{Rubenstein2001} and \citet{Rucinski2004} as follows.
Observations show that two components of these binaries have
nearly equal surface effective temperatures ($T_1=T_2=T$),
while their radius ratio is constrained by Roche geometry,
$R_2/R_1 \approx (M_2/M_1)^{0.46}$ \citep{Kuiper1941}. Their
total surface luminosities are nearly equal to their total
nuclear luminosities \citep[e.g.][]{Webbink2003, Jiang2009}.
Considering the luminosity transfer between two components and
the primaries still in the main sequence \citep{Yakut2005,
Li2008}, we can obtain these luminosities and effective
temperatures of two components of contact binaries, and then
their combined magnitudes when these binaries are
observed as one point.

Figure 9 shows the distribution of contact binaries with mass
ratios $q=0.1, 0.3, 0.5, 0.7, 0.9$ in the color-magnitude
diagram, while the primary masses range from 0.91\,M$_{\rm
\odot}$ to 1.44\,M$_{\rm \odot}$. These contact binaries become
brighter and redder as the mass ratio increases, and this shift
agrees with the results given by \citet{Rucinski2004}. These
contact binaries with different mass ratio can cover the
observed distribution of W UMa contact binaries in M30, and we
find that the observed W UMa contact binary in the blue
sequence corresponds to a smaller mass ratio (e.g. $q < 0.3$)
than the two W UMa contact binaries on the red sequence (e.g.
$q > 0.3$). This suggests that W UMa contact binaries in both
sequences of blue stragglers may be due to their different mass
ratio.

Moreover, we suggest that W UMa contact binaries could
evolve from the red sequence to the blue sequence, when they
evolve into systems with smaller mass ratios due to dynamical
evolution \citep{Li2008}. At present, there are about a dozen W
UMa systems with known mass ratios in globular clusters
\citep[e.g.][]{Kallrath1992, McVean1997, Li2013}, and only half
of them are located in the blue-straggler region. According to
the color-magnitude diagram of NGC 6397 given by
\citet{Kaluzny2006}, the W UMa system V8 seems to be a
blue-sequence blue straggler, which has a low mass ratio
\citep[$q=$0.159,][]{Li2013}. This may be consistent with our
prediction.

Despite a similar origin of binary evolution, W UMa contact
binaries in M30 have different Roche-lobe-filling situations
from other blue-straggler binaries. For other blue-straggler
binaries, those in the red sequence are the semi-detached or
detached BS-WD binaries, while those in the blue sequence are
mainly detached BS-WD binaries. However, W UMa contact binaries
can appear at different sequences in the color-magnitude
diagram for the reason similar to other blue-straggler
binaries. The reason is that the contributions of the
less-massive companions are different when they are observed as
a single star. Therefore, the binary scenario is not
incompatible with the observed W UMa contact binaries in M30.

\section{Conclusions}
In this paper, we explore the possibility that binary evolution
contributes to the formation of blue stragglers in two
sequences in globular cluster M30. Our results show that the
primordial binaries may contribute to the blue sequence of blue
stragglers in M30. Considering W UMa contact binaries observed
in both sequence, this possibility of binary evolution having
contribution to both sequences should not be ruled out. We
suggest that this feature, a blue sequence with a much wider
red sequence, may not be uncommon among globular clusters.
However, the observed fraction of blue sequence cannot be
reproduced by the studies of binary population synthesis with
the initial distribution of field binaries, which suggests that
initial distribution of binaries in globular clusters may be
modified by dynamical interaction or be very different from
that in the field.

\acknowledgments

It is a pleasure to thank an anonymous referee for many
valuable suggestions and comments, which improved the paper
greatly. We thank Professor L. Deng, Professor F. Ferraro,
Professor A. Sills and Dr Y. Xin for the helpful discussions.
This work is supported by the Natural Science Foundation of
China (Nos 11573061, 11521303, 11733008, 11422324, 11773065 and
11661161016), by the Yunnan province (Nos 2017HC018, 2013HA005,
2015FB190).

\appendix

\clearpage

\begin{table}
\begin{minipage}{160mm}
\caption{Sets in different simulations. IMF = initial mass
funcation; $n(q_0)=$ initial mass ratio distribution;  $a =$
the distribution of orbital separation.}
\begin{tabular}{ccccc}
\hline
Set & IMF & $n(q_0)$ & $a$\\
\hline
$1$ & ${\rm MS79}$     & ${\rm Constant}$        & ${\rm Constant}$ \\
$2$ & ${\rm MS79}$     & ${\rm Rising}$          & ${\rm Constant}$ \\
$3$ & ${\rm MS79}$     & ${\rm Uncorrelated}$    & ${\rm Constant}$ \\
$4$ & ${\rm S86}$      & ${\rm Constant}$        & ${\rm Constant}$ \\
$5$ & ${\rm MS79}$     & ${\rm Constant}$        & ${\rm DM91}$ \\
\hline
\end{tabular}
\end{minipage}
\end{table}

\begin{table}
\begin{minipage}{160mm}
\caption{Total number of blue straggler and the fraction of
blue sequence for different simulation sets. IMF = initial mass
funcation; $n(q_0)=$ initial mass ratio distribution;  $a =$
the distribution of orbital separation; $f_{\rm b} =$ the
initial fraction of binaries; $N_{\rm total} = $ the total
numbers of blue stragglers; $N_{\rm blue}/N_{\rm total} = $ the
fraction of blue sequence. }
\begin{tabular}{cccccccc}
\hline
Set & IMF & $n(q_0)$ & $a$ & $f_{\rm b}$ & $N_{\rm all}$ & $N_{\rm blue}/N_{\rm total}$\\
\hline
$1$ & ${\rm MS79}$     & ${\rm Constant}$        & ${\rm Constant}$ & ${\rm 25\%}$ & ${\rm 77}$ & ${\rm 0.65}$\\
$ $ & $          $     & $              $        & $              $ & ${\rm 15\%}$ & ${\rm 46}$ & ${\rm 0.65}$\\
$2$ & ${\rm MS79}$     & ${\rm Rising}$          & ${\rm Constant}$ & ${\rm 25\%}$ & ${\rm 84}$ & ${\rm 0.60}$\\
$ $ & $          $     & $              $        & $              $ & ${\rm 15\%}$ & ${\rm 50}$ & ${\rm 0.60}$\\
$3$ & ${\rm MS79}$     & ${\rm Uncorrelated}$    & ${\rm Constant}$ & ${\rm 25\%}$ & ${\rm 45}$ & ${\rm 0.70}$\\
$ $ & $          $     & $              $        & $              $ & ${\rm 15\%}$ & ${\rm 26}$ & ${\rm 0.70}$\\
$4$ & ${\rm S86}$      & ${\rm Constant}$        & ${\rm Constant}$ & ${\rm 25\%}$ & ${\rm 99}$ & ${\rm 0.64}$\\
$ $ & $          $     & $              $        & $              $ & ${\rm 15\%}$ & ${\rm 60}$ & ${\rm 0.64}$\\
$5$ & ${\rm MS79}$     & ${\rm Constant}$        & ${\rm DM91}$     & ${\rm 25\%}$ & ${\rm 18}$ & ${\rm 0.67}$\\
$ $ & $          $     & $              $        & $              $ & ${\rm 15\%}$ & ${\rm 10}$ & ${\rm 0.67}$\\
\hline
\end{tabular}
\end{minipage}
\end{table}

\begin{figure}
\epsscale{0.5} \plotone{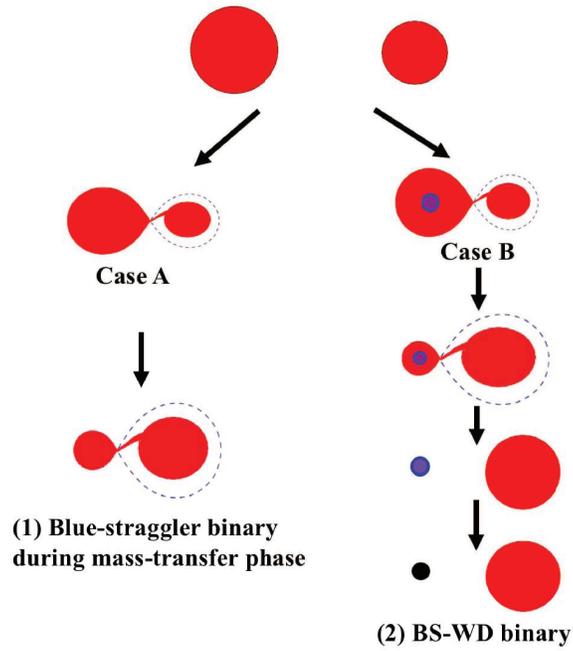} \caption{Schematic view of
the evolutionary paths of binaries
that produce two kinds of blue-straggler binaries: (1) Case A binary evolution may
produce a blue-straggler binary that are experiencing mass transfer; (2) Case B
binary evolution may produce a blue-straggler binary that has finished mass
transfer, which has a blue
straggler orbiting a white dwarf (the BS-WD binaries). \label{fig0}}
\end{figure}

\begin{figure}
\epsscale{0.9} \plotone{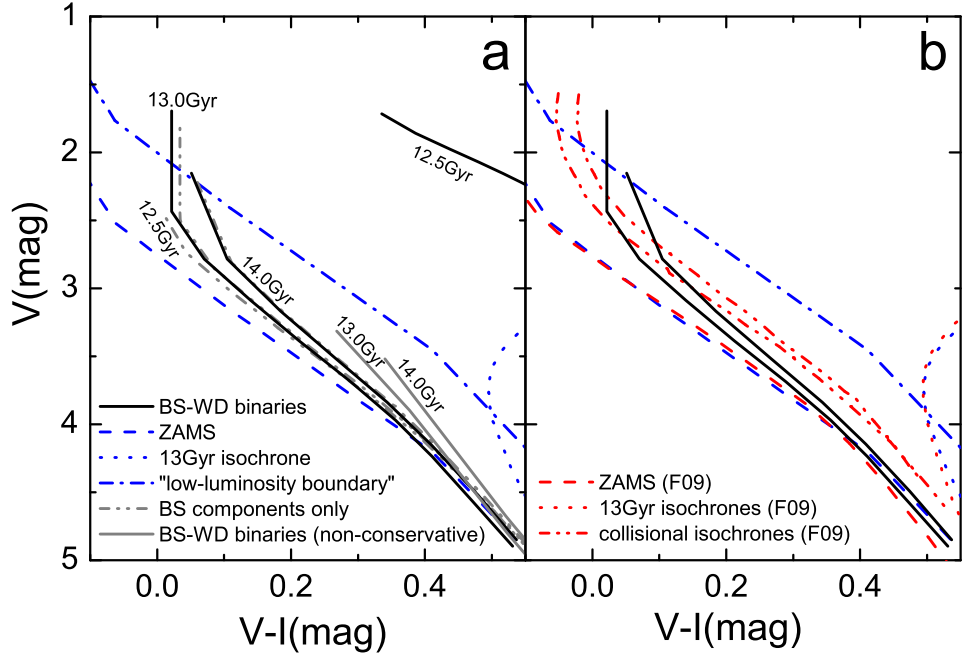} \caption{The locations of
BS-WD binaries roughly estimated based on case
B binary evolution ($M_{10}=0.8$\,M$_{\rm \odot}$) in conservative mass-transfer cases.
In panel (a), black solid lines show their locations
at the end of mass
transfer (12.5\,Gyr), at the age of M30 (13.0\,Gyr) and
at the subsequent evolution (14.0\,Gyr). The
gray dash-dot-dot lines show the blue-straggler components,
while the gray solid lines present the BS-WD binaries in
non-conservative mass-transfer cases (50\% of the transferred
mass is assumed to be lost). Blue dotted, dashed and dash-dot lines correspond to the
single-star isochrone of 13\,Gyr, the ZAMS and the
``low-luminosity boundary" (0.75\,mag brighter than the ZAMS),
respectively. Panel (b) shows the comparison between the
BS-WD binaries and collisional
isochrones (red dash-dot-dot lines) corresponding to ages of 1 and 2\,Gyr
given by \citet{Sills2009},
which agree with the
observed blue sequence as shown by \citet{Ferraro2009}.
The collisional isochrones have been
transferred into the absolute plane using a distance modulus of
$(m-M)_{\rm v} = 15.04$\,mag and a reddening of $E(V-I) =
0.112$\,mag (see the text).
\label{fig0}}
\end{figure}

\begin{figure}
\epsscale{0.9} \plotone{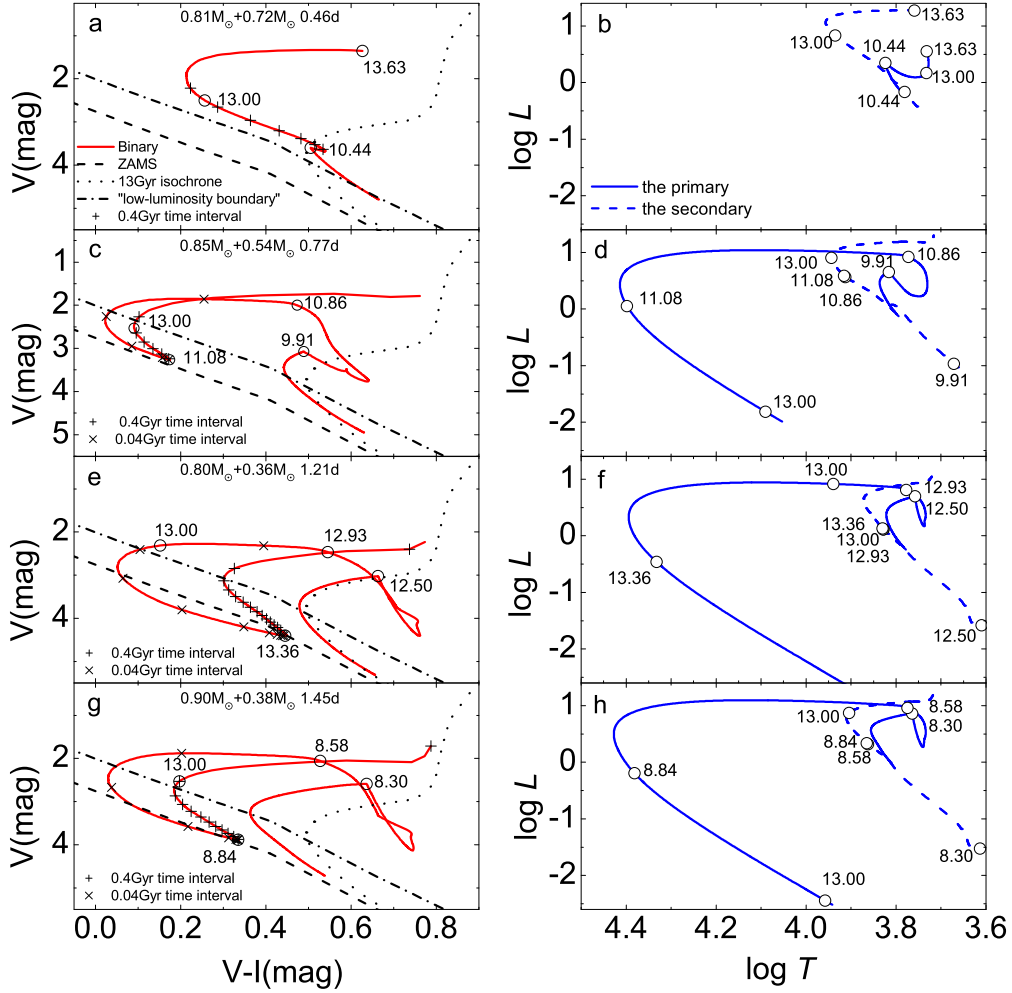}\caption[]{Four examples
of binary evolution for the formation of blue stragglers in two
sequences at the age of M30 (13\,Gyr): (1) red sequence: the
blue-straggler binary experiencing mass transfer; (2) blue
sequence: BS-WD binary while the white dwarf is cooled down
and has little contribution; (3) red sequence: BS-WD binary while the
white dwarf is bright and hot; (4) red sequence: BS-WD binary
while the blue straggler has evolved away from the main sequence.
Red solid lines show the combined
evolutionary tracks of the binary systems in left Panels, while
blue solid lines for the primaries and blue dashed lines for
the secondaries in right Panels. The initial binary parameters
are also given in left panels. The numbers along the tracks
give the ages (in Gyr) of selected phases (black open circles):
the begin of mass transfer, the age of M30 and the final model
for the first example; the begin of mass transfer, the end of
mass transfer, the largest $V$-band magnitude and the age of
M30 other examples. Dotted, dashed and dash-dot lines
correspond to the single-star isochrone of 13\,Gyr, the ZAMS
and the ``low-luminosity boundary", respectively. In panel (a)
and (c), the time intervals between two pluses are 0.4\,Gyr,
and that between two crosses are 0.04\,Gyr. } \label{fig1}
\end{figure}


\begin{figure}
\epsscale{0.9} \plotone{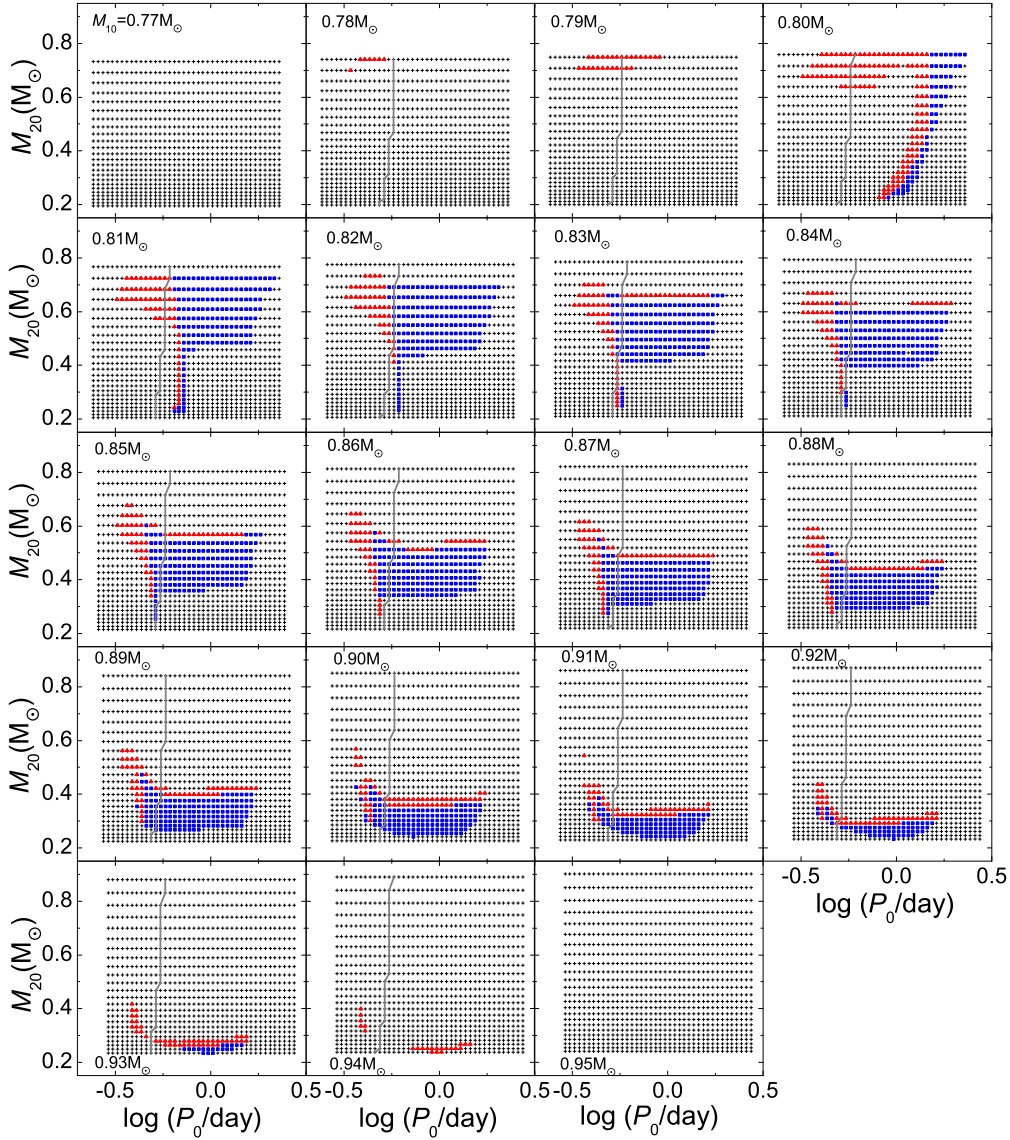} \caption{Initial binary parameters
of blue-straggler binaries in the two
sequences. Blue squares and red triangles are initial binaries leading to blue stragglers in the blue
and the red sequences, respectively. Black `plus' signs show
systems that are unable to form blue stragglers at 13\,Gyr.
Each panel is for a primary with initial mass
($M_{10}$) as indicated, while $M_{20}$ and $P_0$ are the initial mass of
the secondaries and initial orbital period. The gray lines show
the boundary between case A and case B binary evolution. \label{fig2}}
\end{figure}


\begin{figure}
\epsscale{0.9} \plotone{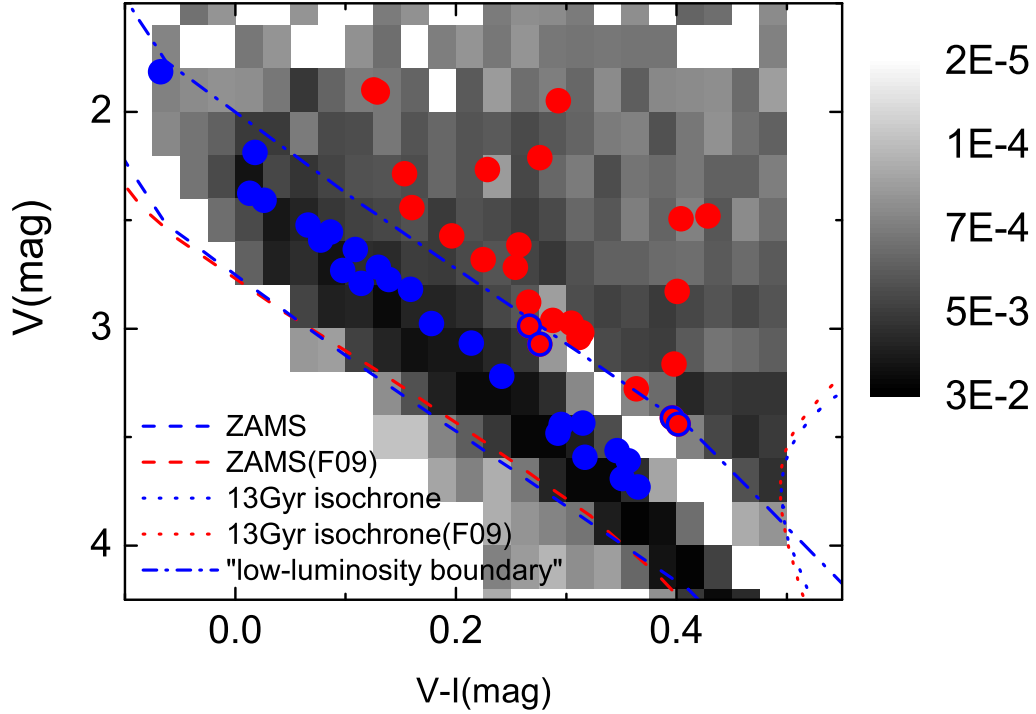} \caption{Monte Carlo simulation of binary-origin blue
stragglers with an age of 13\,Gyr in the color-magnitude diagram and comparison
with the observed blue stragglers in M30. It shows the
density plot for the fraction of blue stragglers for simulations set 1.
Blue dotted, blue dashed and blue dash-dot lines
indicate the 13Gyr isochrone of single stars, the ZAMS and the
``low-luminosity boundary" as shown in Figure 3a.
Red and blue points are observed blue stragglers in the
red and blue sequences in
M30 \citep{Ferraro2009,Xin2015}. Four of the red points below
the ``low-luminosity boundary" are marked with blue circles.
These observed blue stragglers have been transferred into the
absolute plane using the distance modulus of $(m-M)_{\rm v} =
15.04$\,mag and the reddening of $E(V-I) = 0.112$\,mag, which
are obtained in Figure 2b.
These binary-origin blue stragglers can explain the
distribution of blue stragglers observed in M30, include the
presence of the blue sequence. A blue-straggler sequence is
located between the ZAMS and the ``low-luminosity boundary",
similar to the observed blue sequence (about half a magnitude
brighter than the ZAMS).  \label{fig3}}
\end{figure}

\begin{figure}
\epsscale{0.9} \plotone{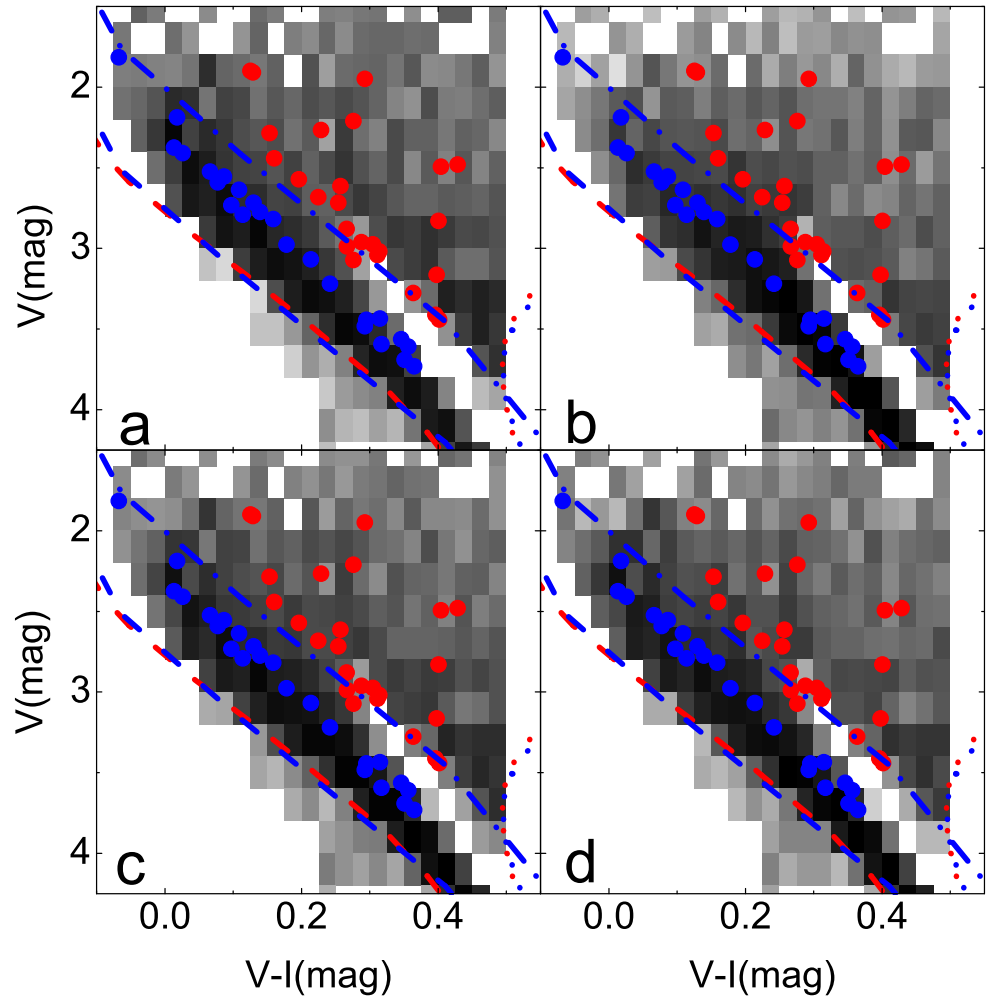} \caption{Similar to Fig.3,
but for simulation sets 2-5 (Panel a-d). \label{fig4}}
\end{figure}

\begin{figure}
\epsscale{0.9} \plotone{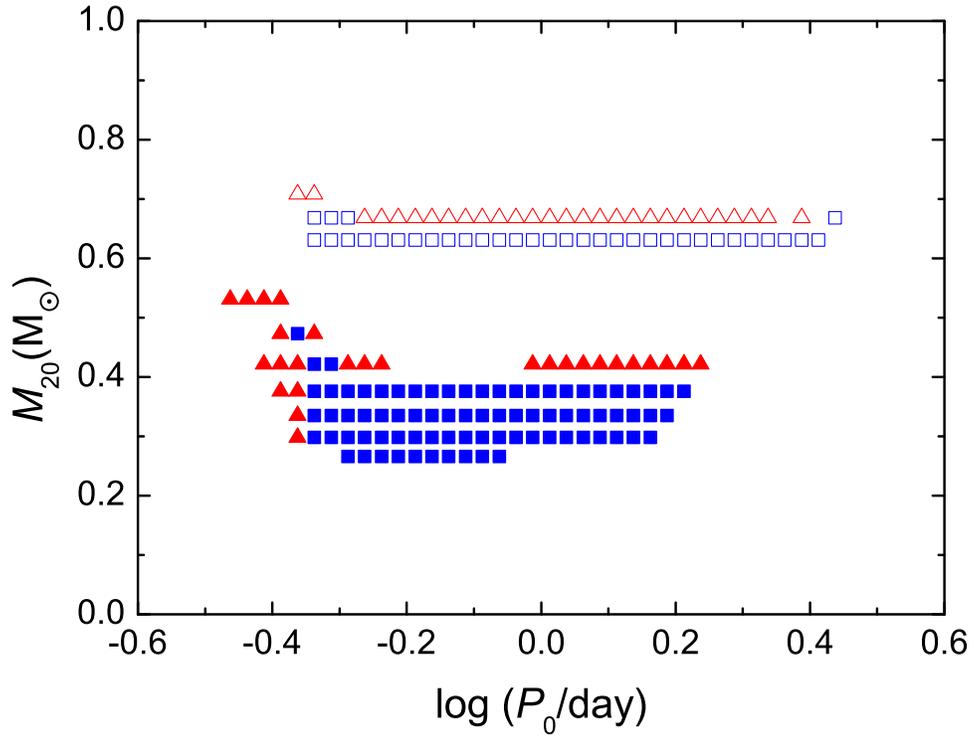} \caption{Comparison
the progenitor region of two sequences in the conservative and
non-conservative mass-transfer cases for binaries with a
primary of 0.89M$_{\rm \odot}$. Blue squares and red triangles represent the
blue and red sequences in the conservative cases, while those open squares
and open triangles are for the non-conservative cases. \label{fig5}}
\end{figure}



\begin{figure}
\epsscale{1.0} \plotone{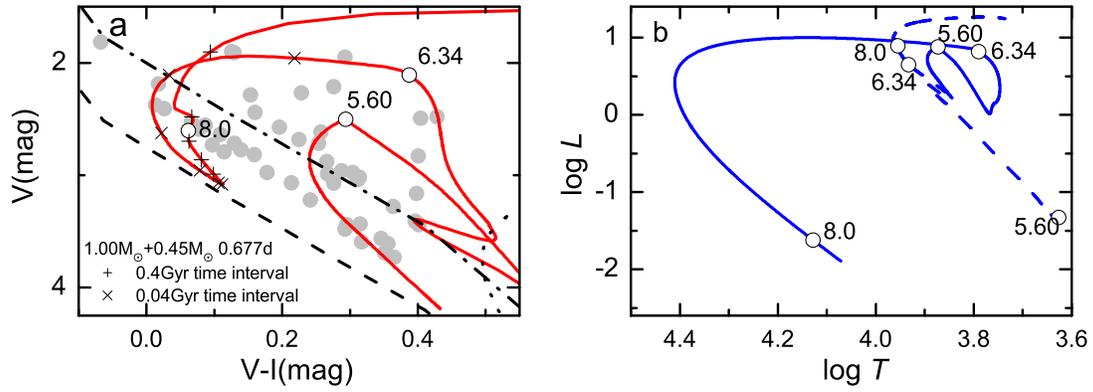} \caption{Similar to Fig. 3,
but for the binary evolution that produces a BS-WD binary in the blue sequence
at the age of 8\,Gyr. \label{fig7}}
\end{figure}

\begin{figure}
\epsscale{0.9} \plotone{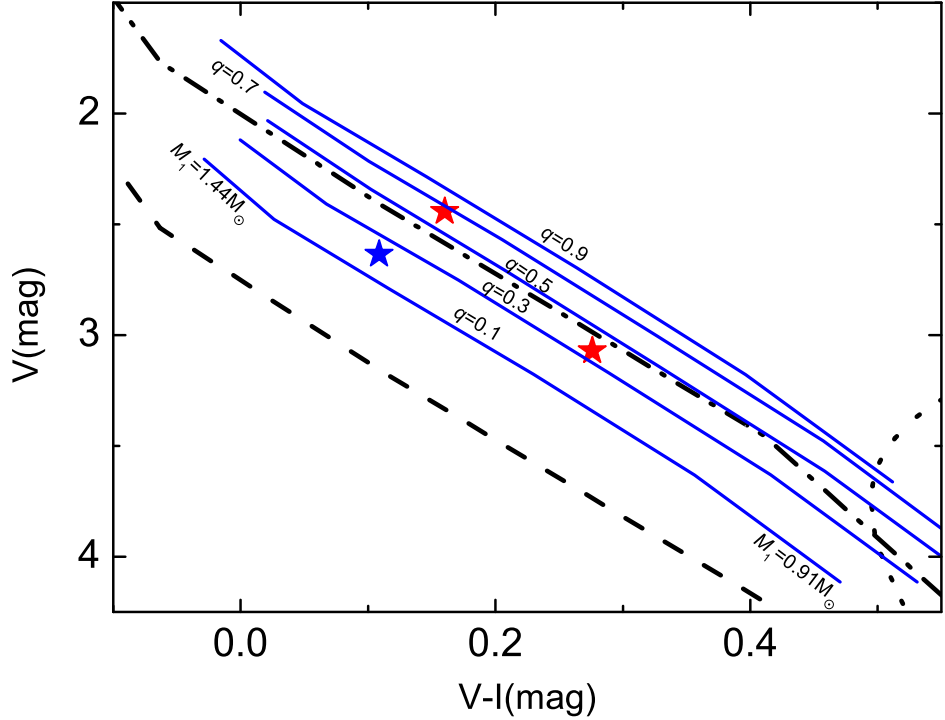} \caption{Comparison with observed
W UMa contact binaries
with different mass-ratio models in the color-magnitude diagram.
The dotted, dashed and dash-dot lines indicate the
13Gyr isochrone, the ZAMS and the
``low-luminosity boundary" as shown in Figure 3a.
Three filled asterisks are W UMa
contact binaries observed in the blue-straggler region of
M30 \citep{Ferraro2009}.
Each solid
blue line show the locations of contact binaries with the same
mass ratio but with different primary mass, and the
corresponding mass ratios of these lines, from the left to
right, are $q=0.1, 0.3, 0.5, 0.7, 0.9$, respectively. The
primary masses along the blue lines range from 0.91\,M$_{\rm
\odot}$ to 1.44\,M$_{\rm \odot}$ at equal intervals in log\,$M$
(log\,$M = -0.04, 0, 0.04, 0.08, 0.12, 0.16$). \label{fig6}}
\end{figure}

\clearpage

\end{document}